\documentclass[12pt]{article}
\usepackage{multirow}

\begin{document}
\def\qq{\langle \bar q q \rangle}
\def\uu{\langle \bar u u \rangle}
\def\dd{\langle \bar d d \rangle}
\def\sp{\langle \bar s s \rangle}
\def\GG{\langle g_s^2 G^2 \rangle}
\def\Tr{\mbox{Tr}}
\def\figt#1#2#3{
        \begin{figure}
        $\left. \right.$
        \vspace*{-2cm}
        \begin{center}
        \includegraphics[width=10cm]{#1}
        \end{center}
        \vspace*{-0.2cm}
        \caption{#3}
        \label{#2}
        \end{figure}
	}
	
\def\figb#1#2#3{
        \begin{figure}
        $\left. \right.$
        \vspace*{-1cm}
        \begin{center}
        \includegraphics[width=10cm]{#1}
        \end{center}
        \vspace*{-0.2cm}
        \caption{#3}
        \label{#2}
        \end{figure}
                }
\def\ph{\phantom{-}}

\def\ds{\displaystyle}
\def\beq{\begin{equation}}
\def\eeq{\end{equation}}
\def\bea{\begin{eqnarray}}
\def\eea{\end{eqnarray}}
\def\beeq{\begin{eqnarray}}
\def\eeeq{\end{eqnarray}}
\def\ve{\vert}
\def\vel{\left|}
\def\ver{\right|}
\def\nnb{\nonumber}
\def\ga{\left(}
\def\dr{\right)}
\def\aga{\left\{}
\def\adr{\right\}}
\def\lla{\left<}
\def\rra{\right>}
\def\rar{\rightarrow}
\def\lrar{\leftrightarrow}  
\def\nnb{\nonumber}
\def\la{\langle}
\def\ra{\rangle}
\def\ba{\begin{array}}
\def\ea{\end{array}}
\def\tr{\mbox{Tr}}
\def\ssp{{\Sigma^{*+}}}
\def\sso{{\Sigma^{*0}}}
\def\ssm{{\Sigma^{*-}}}
\def\xis0{{\Xi^{*0}}}
\def\xism{{\Xi^{*-}}}
\def\qs{\la \bar s s \ra}
\def\qu{\la \bar u u \ra}
\def\qd{\la \bar d d \ra}
\def\qq{\la \bar q q \ra}
\def\gGgG{\la g^2 G^2 \ra}
\def\q{\gamma_5 \not\!q}
\def\x{\gamma_5 \not\!x}
\def\g5{\gamma_5}
\def\sb{S_Q^{cf}}
\def\sd{S_d^{be}}
\def\su{S_u^{ad}}
\def\sbp{{S}_Q^{'cf}}
\def\sdp{{S}_d^{'be}}
\def\sup{{S}_u^{'ad}}
\def\ssp{{S}_s^{'??}}

\def\sig{\sigma_{\mu \nu} \gamma_5 p^\mu q^\nu}
\def\fo{f_0(\frac{s_0}{M^2})}
\def\ffi{f_1(\frac{s_0}{M^2})}
\def\fii{f_2(\frac{s_0}{M^2})}
\def\O{{\cal O}}
\def\sl{{\Sigma^0 \Lambda}}
\def\es{\!\!\! &=& \!\!\!}
\def\ap{\!\!\! &\approx& \!\!\!}
\def\ar{&+& \!\!\!}
\def\ek{&-& \!\!\!}
\def\kek{\!\!\!&-& \!\!\!}
\def\cp{&\times& \!\!\!}
\def\se{\!\!\! &\simeq& \!\!\!}
\def\eqv{&\equiv& \!\!\!}
\def\kpm{&\pm& \!\!\!}
\def\kmp{&\mp& \!\!\!}
\def\mcdot{\!\cdot\!}
\def\erar{&\rightarrow&}

% .........................................................

\def\simlt{\stackrel{<}{{}_\sim}}
\def\simgt{\stackrel{>}{{}_\sim}}

% .........................................................

\title{
         {\Large
                 {\bf
Vector meson dominance and radiative decays of heavy spin--3/2 baryons to heavy
spin--1/2 baryons
                 }
         }
      }

\author{\vspace{1cm}\\
{\small T. M. Aliev$^a$ \thanks
{e-mail: taliev@metu.edu.tr}~\footnote{permanent address:Institute
of Physics,Baku,Azerbaijan}\,\,,
M. Savc{\i}$^a$ \thanks
{e-mail: savci@metu.edu.tr} \,\,,
V. S. Zamiralov$^b$ \thanks  
{e-mail: zamir@depni.sinp.msu.ru}} \\
{\small (a) Physics Department, Middle East Technical University,
06531 Ankara, Turkey} \\
{\small (b) Institute of Nuclear Physics, M. V. Lomonosov MSU, Moscow,
Russia} }

\date{}
\begin{titlepage}
\maketitle
\thispagestyle{empty} 

\begin{abstract}
Using the calculated values of the strong coupling constants of the heavy
sextet spin--3/2 baryons to sextet and antitriplet heavy spin--1/2 baryons
with light mesons within the light cone QCD sum rules method, and vector
meson dominance assumption, the radiative decay widths are calculated.
These widths are compared with the ``direct" radiative decay widths 
predicted in the framework of the light cone QCD sum rules.
\end{abstract}

~~~PACS number(s): 11.55.Hx, 13.30.Eg, 14.20.Mr, 14.20.Lq
\end{titlepage}

%\section{Introduction}

Heavy baryons with a single heavy quark are quite promising for testing the
predictions of the heavy symmetry and quark models (for a recent review see 
\cite{Rvmd01}).
Last few years have been very successful on heavy baryon spectroscopy.
Practically all $1/2^+$ and $3/2^+$ ground states with a single heavy quark
have been discovered in the experiments \cite{Rvmd02}.

With the operation of LHC, new possibility is opened for a comprehensive study
of the properties of the heavy flavor hadrons as well as their
electromagnetic, weak and strong decays \cite{Rvmd03}. The coupling constant
of the heavy baryons with light mesons is the main ingredient for describing
the strong decays. The strong coupling
constants of the heavy sextet 3/2 baryons $B^*$  with the spin--1/2 sextet and
antitriplet heavy baryons $B$ with light vector mesons $V$ within the light cone QCD sum
rules (LCSR) are analyzed in \cite{Rvmd04}.

In the present note we study the electromagnetic decays of heavy flavored
hadrons assuming vector--meson dominance model (VDM) by using the results 
for the strong coupling constants of light vector mesons with heavy hadrons
\cite{Rvmd04}.
Important feature of these decays is also the fact that although electromagnetic interaction
involves the constant of the fine structure $\alpha$ they are not suppressed
by the phase space, as is the case for pion transitions. Moreover radiative decay
for some 3/2 baryons would be the main decay mode.
 The secondary aim of this note is to find an answer to the
question how VDM works for transitions of heavy hadrons.

Let us start our discussion on
the heavy baryon decays in the unitary symmetry.
%We start with reminding some points from the unitary symmetry model with 5  %flavors.
The 3/2 $B^*$ $\rightarrow$ 1/2 $B$ electromagnetic baryon current 
in the unitary symmetry model with 5 flavors can be written as
\bea
\label{Ro4}
J^{3/2\rightarrow 1/2~el}_{\mu}=
%\frac23 J^1_{\mu ~1}-
%\frac13 J^2_{\mu ~2}-\frac13 J^3_{\mu ~3}+\frac23 J^4_{\mu ~4}
%-\frac13 J^5_{\mu ~5}=\nnb\\
\sum_{k=1}^5 e_k J^k_{\mu ~k}~,
%\nnb\\
\qquad \qquad J^{\tau}_{\mu~\delta}=
\epsilon_{\beta\gamma\rho \delta\eta}{\cal {O}}_{\mu\nu}
\bar B^{\beta\gamma\eta}_{\alpha}B^{*\nu~\alpha\rho\tau}~,
\eea
where $\mu,\nu$=0,1,2,3 are the Lorenz indices, and
all other indices run from 1 to 5, and $e_k$ means $k$'th flavor 
electric charge $k=1,2,3,4,5$ which is, naturally,
to associate with the electric charge of the quarks $u,d,s,c,b$, 
correspondingly;
and we do not specify the nature of ${\cal {O}}_{\mu\nu}$ for a moment.

The baryon $1/2^+$ wave function in terms
of the quarks is
\bea
\sqrt{6}B^{\alpha}_{\beta \gamma\tau\uparrow}
=\epsilon_{\beta \gamma \eta \rho\tau} 
\{q^{\alpha}_{\uparrow},q^{\eta}_{\uparrow} \}q^{\rho}_{\downarrow}~,
\hspace{4cm}
\eea
where $u=q^{1},d=q^{2},s=q^{3},c=q^{4},b=q^{5}$ (subscripts 1(2) or 
${\uparrow}$(${\downarrow}$) mean spin up (down)). 
%For completeness we present baryon wave functions in Appendix A.

Neglecting baryon currents related with heavy quarks Eq.(1) can be written as
\bea
\label{Ro44}
J^{3/2\rightarrow 1/2~el}_{\mu}=
\frac12 (e_u-e_d)( J^1_{\mu ~1}-J^2_{\mu ~2})+\frac12 (e_u+e_d)( J^1_{\mu ~1}+J^2_{\mu ~2})+
e_s J^3_{\mu ~3}~,
\eea
The baryon currents have the same quantum numbers of $\rho^0$, $\omega$ and $\phi$ mesons, respectively.
It is easy now to express all the electromagnetic quantities through the VDM hypothesis in terms of
the $B^*BV$ couplings, where $V=\rho^0, \omega,\phi$.

Now we turn to the explicit forms of ${\cal {O}}_{\mu\nu}$. 
%Since $q^2=0$ and $q\varepsilon=0$ for photons, the last term in Eq.
%(\ref{eali09}) can be dropped. 
Using the gauge invariance the amplitude $B^* \to B \gamma$ is
parametrized in terms of form factors as follows \cite{Rvmd05}:
\bea
\label{eali12}
\lla B_Q(p) \gamma(q)  \ve B^*(p+q) \rra \es \bar{u} (p) \Big\{ g_1 \left(
q_\mu \rlap/{\varepsilon} - \varepsilon_\mu \rlap/{q} \right) + g_2 \left[ \left( P
\varepsilon \right) q_\mu - \left(P q\right) \varepsilon_\mu \right]\nnb\\
\ar g_3^V \left[ \left( q \varepsilon \right) q_\mu - q^2 \varepsilon_\mu \right]
\Big\} \gamma_5 u^\mu (p+q)~,
\eea
where $u^\mu$ is the Rarita Schwinger spinor, $\varepsilon_\mu$ is the photon
polarization 4--vector, $q_\mu$ is its momentum, $p$ being momentum of the 1/2 baryon and $P=2p+q$.
Obviously for real photons last term in the Eq.(\ref{eali12}) is equal to zero.
%\section{Vector-dominance model for heavy baryon decays}

The VDM implies that  $B^*B\gamma$ vertex can be obtained from $B^*BV$ vertex
by converting vector meson $V$ to photon, i.e., $B^*BV \to B^*B\gamma$. 
The corresponding strong  $B^*BV$ vertex is parametrized in terms of the form factors
similar to the   $B^*B\gamma$ vertex \cite{Rvmd06}:
\bea
\label{eali09}
\lla B_Q(p) V(q)  \ve B^*(p+q) \rra \es \bar{u} (p) \Big\{ g_1^V \left(
q_\mu \rlap/{\eta}^V - \eta_\mu \rlap/{q} \right) + g_2^V \left[ \left( P
\eta^V \right) q_\mu - \left(P q\right) \eta_\mu^V \right] \nnb \\
\ar g_3^V \left[ \left( q \eta^V \right) q_\mu - q^2 \eta_\mu^V \right]
\Big\} \gamma_5 u^\mu (p+q)~,
\eea
where now $\eta_\mu^V$ is the vector
polarization of light vector meson with the momentum $q_\mu$.
We disregard the form factor $g_3$ as it corresponds to
longitudinal polarization not presented in photon.
In this chain it is necessary to go from $q^2=m_V^2$ to
$q^2=0$ and make the replacement,
\bea
\label{eali10}
\eta_\mu^V = {e \over g_V} e_\mu~.
\eea
Obviously, when $e_\mu \to q_\mu$, the $B^*B\gamma$ amplitude
should vanish as is required by the gauge invariance. 

Putting Eq. (\ref{eali10}) in Eq.(\ref{eali09}), we get
\bea
\label{eali11}
\lla B_Q(p) \gamma(q)  \ve B^*(p+q) \rra \es \sum_{V=\rho,\omega,\phi}
{e\over g_V} \bar{u} (p) \Big\{ g_1^V \left( 
q_\mu \rlap/{\varepsilon} - \varepsilon_\mu \rlap/{q} \right) \nnb \\
\ar g_2^V \left[ \left( P   
\varepsilon \right) q_\mu - \left(P q\right) \varepsilon_\mu \right]
\Big\} \gamma_5 u^\mu (p+q)
~.
\eea

Comparing Eqs. (\ref{eali11}) and (\ref{eali12}), we obtain the usual
relation among the form factors of $B^*BV$ and $B^*B\gamma$ vertices,
\bea
\label{nolabel}
g_i = \sum_{V=\rho,\omega,\phi} g_i^V {1\over g_V}~.\nnb     
\eea
It should be noted here that we neglect the possible contributions coming
from $J/\Psi$ and $\Upsilon$ mesons, since $B^*BV$ vertex is proportional to  
$m_{J/\Psi,(\Upsilon)}^{-3/2}$ in the heavy quark limit 
(This is not a general rule, as for example in quark model approach charm quark contribution
is taken comparable with that of strange quark \cite{Rvmd07}).
In numerical analysis we
use the experimental values
$g_\rho=5.05$, $g_\omega=17.02$ and $g_\phi=-12.89$ \cite{Rvmd02}.

From experimental point of view multipole form factors are
much more suitable than the form factors $g_1$ and $g_2$ themselves. In our
case the multipole form factors are the magnetic dipole $G_M$ and electric
quadrupole $G_E$ form factors. Relations among
$G_M$, $G_E$ and $g_1$ and $g_2$ at $q^2=0$ can be found in \cite{Rvmd06},
whose forms are as follows:
\bea
\label{egm12}
G_M \es \left( 3 m_{B^*} + m_B \right) {m_B \over 3m_{B^*} } g_1 + \left( m_{B^*}
- m_B \right) m_B {g_2 \over 3}~, \nnb \\
G_E \es \left( m_{B^*} - m_B \right) {m_B \over 3 m_B^*} \left(g_1 + m_B^*
g_2\right)~.
\eea
Using these relations it is straightforward to get the 
width of the radiative decay $B^*_Q \rightarrow
B_Q \gamma $, which can be written as:
\bea
\label{Ro11}
\Gamma_{B^*_Q \rightarrow B_Q \gamma }\es \frac{3\alpha}{4}\frac{k_\gamma^3}
{m^2_{B_Q}} (3G_E^2+G_M^2)~,
%\eea
%where 
%\bea
%\label{nolabel}
\qquad \qquad k_\gamma = {m_{B^*}^2 - m_B^2 \over 2 m_{B^*}}~. \nnb
\eea
%is the photon energy.
%\section{Numerical analysis and discussion of results}

Now we implement the VDM hypothesis through the Eq.(\ref{eali10}) into the
LCSR and calculate
the $G_M$'s and $G_E$'s. 
The results of our calculations we put into the tables together with the results of some previous works.

We expect that similar to the case of the decuplet--octet radiative decays
$G_E$'s are suppressed in comparison with $G_M$'s. Indeed,
this is confirmed by explicit calculations.
First let us try to understand the possible tendency of the VDM calculations together
with the QCD results on the examples of measured quantities,
$
\Gamma(\Delta^+\to p\gamma)= 660\pm 50~keV
$
\cite{Rvmd02} and 
$
\Gamma(\Sigma^{*0}\to \Lambda\gamma)= 445\pm 80~keV
$
\cite{Rvmd08}.
Both decays in VDM can proceed only through $\rho \to \gamma$ conversion.
So we take $g_1^{\Delta^+\to p\rho^0}$=$\sqrt2 g_1^{\Delta^0\to
p\rho^-}$=$\sqrt{2} (9.1 \pm 2.9)$
and $g_1^{\Sigma^{*0}\to \Lambda\rho^0}$=$g_1^{\Sigma^{*-}\to \Lambda\rho^-}$=$-10\pm 3$ from the Table II of
\cite{Rvmd09}.
With $g_\rho$=5.05 we get $g_1^{VDM}(\Delta^+\to p\gamma)$=2.26, where from
$G_M^{VDM}(\Delta^+\to p\gamma)\sim 1.18 g_1=2.51$
 and
$\Gamma^{VDM}(\Delta^+\to p\gamma)$=860 keV while
$g_1^{VDM}(\Sigma^{*0}\to \Lambda\gamma)$=2.0 and $\Gamma^{VDM}(\Sigma^{*0}\to \Lambda\gamma)$=590 keV.
So for these decays VDM combined with the QCD sum rules has the tendency to oversize data by a factor
1.5--2.0.

Keeping this in mind we return now to heavy baryons decays.
In Table 1 we present the values of the magnetic dipole $G_M$ and 
the electric quadrupole form factors at $q^2=0$, which are obtained within VDM.
In this Table, for a comparison, we also present the values of these form
factors as calculated from direct $B_{Q^*} \to B \gamma$ decay in the
framework of LCSR \cite{Rvmd06}.
The radiative decays of the heavy sextet spin--3/2 baryons to the
sextet and antitriplet spin--1/2
heavy baryons are calculated in this work using the most general form of
the interpolating current  without
using VDM hypothesis.

Using the values of the couplings $G_M$ and $G_E$ presented in Table 1 and
Eq. (\ref{egm12}), we calculate the widths of the radiative decays whose
values are presented in Table 2.

In Table 2 we also present the predictions of the works \cite{Rvmd10},\cite{Rvmd11}
on radiative decays, where VDM hypothesis has been used in
estimating radiative decays of the heavy spin--1/2 baryons 
in the framework of the LCSR approach with the Ioffe interpolating current,
as well as results of several preceding works on a subject
\cite{Rvmd12}--\cite{Rvmd16}.

It follows from Table 2 that, there are considerable differences
between the VDM predictions and the direct
radiative decay ones \cite{Rvmd06}, for all channels.
This difference can be explained as follows.
The decay width is very sensitive to the mass difference of the heavy flavored baryons $B^*$ and $B$.
If the mass of one of the baryons changes even at mili--digit level, the
decay width can change several times.
% The baryon mass we use in the present 
%work is slightly different compared to the one used in \cite{Rvmd06}, which
%causes this big difference in the predictions. 
The baryon masses are taken
from PDG, while in \cite{Rvmd06} for the masses the mass sum rules were used. This is the cause of the
large difference in the predictions for the widths, while there is rather a reasonable agreement for the
values of $G_{M,E}$.

Our results are also different from those predicted in \cite{Rvmd10,Rvmd11}, where VDM has also been
used. This discrepancy in the results is mainly due to the difference of 
the values of the residues of spin--1/2 heavy baryons, 
as well as to the different masses used for the heavy baryons.
 
In conclusion, we estimate the widths of  radiative decays of the heavy 
spin--3/2 baryons to spin--1/2
heavy baryons using the values of strong coupling constants for the
$B^*_Q  B_Q V $ vertices where
$V$ is a light vector meson, and VDM. It is found that VDM works reasonably well for the
heavy--baryon systems.
The obtained results on the radiative widths are compared with the
direct calculation of these decay widths as is predicted
by LCSR method.

\newpage

\begin{table}[tbh]

\renewcommand{\arraystretch}{1.3}
\addtolength{\arraycolsep}{-0.5pt}
\small
$$
\begin{array}{|l|c|c|c|c|c|}
%\begin{array}{|l|c|c|c|c|c|c|c|}
\hline \hline
 \multirow{2}{*}{~~~~~~Channel}                           &
~~~G_E~~~  &
~~~G_M~~~  &
~~~G_E~~~  &
~~~G_M~~~  &
~~~G_M^*~~~    \\
                                                    &
\mbox{VDM}           & \mbox{VDM}           &  \cite{Rvmd06}        & \cite{Rvmd06}       & \cite{Rvmd10},\cite{Rvmd11}        \\ \hline
~~~\Xi^{\ast 0}_b     \rar \Xi^{\prime 0}_b \gamma~~~   & \ph 0.017    & \ph 2.93  & \mbox{---} &  \mbox{---}     & \ph 3.24   \\
~~~\Xi^{\ast -}_b     \rar \Xi^{\prime -}_b \gamma      &    -0.021    &    -4.63  & \mbox{---} &  \mbox{---}     &    -3.86   \\
~~~\Sigma^{\ast +}_b  \rar \Sigma^+_b       \gamma      & \ph 0.016    & \ph 8.35  & \ph 0.026     & \ph 4.2      & \ph 8.85   \\
~~~\Sigma^{\ast 0}_b  \rar \Sigma^0_b       \gamma      & \ph 0.003    & \ph 1.73  & \ph 0.005     & \ph 1.0      & \ph 2.22   \\
~~~\Sigma^{\ast -}_b  \rar \Sigma^-_b       \gamma      &    -0.009    &    -4.53  &    -0.016     &    -2.1      & \ph 4.42   \\     
~~~\Sigma^{\ast 0}_b  \rar \Lambda^0_b      \gamma      & \ph 0.164    & \ph 11.8  & \ph 0.075     & \ph 7.3      & \mbox{---} \\
~~~\Xi^{\ast 0}_b     \rar \Xi^0_b          \gamma      &    -0.185    &    -12.01 &    -0.085     &    -8.5      & \mbox{---} \\ 
~~~\Xi^{\ast -}_b     \rar \Xi^-_b          \gamma      &    -0.025    &    -1.09  &    -0.011     &    -0.9      & \mbox{---} \\
~~~\Omega^{\ast -}_b  \rar \Omega^-_b       \gamma      &    -0.034    &    -4.52  & \mbox{---}    & \mbox{---}   & \mbox{---} \\
\hline \hline
~~~\Xi^{\ast +}_c     \rar \Xi^{\prime +}_c \gamma      & \ph 0.019    & \ph 1.33  & \mbox{---}    & \mbox{---}   &\ph 1.86    \\
~~~\Xi^{\ast 0}_c     \rar \Xi^{\prime 0}_c \gamma      &    -0.026    &    -2.20  & \mbox{---}    & \mbox{---}   &\ph 2.22    \\
~~~\Sigma^{\ast ++}_c \rar \Sigma^{++}_c    \gamma      & \ph 0.037    & \ph 3.68  & \ph 0.030     & \ph 2.8      &\ph 5.22    \\
~~~\Sigma^{\ast +}_c  \rar \Sigma^+_c       \gamma      & \ph 0.009    & \ph 0.84  &    -0.003     & \ph 0.5      &\ph 1.31    \\
~~~\Sigma^{\ast 0}_c  \rar \Sigma^0_c       \gamma      &    -0.020    &    -2.00  & \ph 0.014     & \ph 1.2      &\ph 2.60    \\     
~~~\Sigma^{\ast +}_c  \rar \Lambda^+_c      \gamma      & \ph 0.192    & \ph 5.70  & \ph 0.060     & \ph 3.8      &\mbox{---}  \\
~~~\Xi^{\ast +}_c     \rar \Xi^+_c          \gamma      &    -0.147    &    -5.91  &    -0.075     &    -4.0      &\mbox{---}  \\ 
~~~\Xi^{\ast 0}_c     \rar \Xi^0_c          \gamma      &    -0.013    &    -0.55  &    -0.007     &    -0.45     &\mbox{---}  \\
~~~\Omega^{\ast 0}_c  \rar \Omega^0_c       \gamma      &    -0.026    &    -2.17  & \mbox{---}    & \mbox{---}   &\ph 2.14    \\
\hline \hline

\end{array}
$$
\caption{The values of the electric quadrupole $G_E$ and the magnetic dipole
$G_M$ form factors at $q^2=0$. $G_M^*$  are calculated from $g_1$'s of \cite{Rvmd10},\cite{Rvmd11} neglecting $g_2$'s.}

\renewcommand{\arraystretch}{1}
\addtolength{\arraycolsep}{-1.0pt}
\end{table}

\newpage

\begin{table}[tbh]

\renewcommand{\arraystretch}{1.3}
\addtolength{\arraycolsep}{-0.5pt}
\small
$$
\begin{array}{|l|c|c|c|c|}
%\begin{array}{|l|c|c|c|c|c|c|c|}
\hline \hline
 \multirow{2}{*}{~~~~~~Channel}                           &
~~~\Gamma~(keV)~~~  &
~~~\Gamma~(keV)~~~  &
~~~\Gamma~(keV)~~~  &
~~~\Gamma~(keV)~~~ \\
                                                    &
\mbox{VDM}           &  \cite{Rvmd06}           &  \cite{Rvmd10}, \cite{Rvmd11}    &    \\ \hline
~~~\Xi^{\ast 0}_b     \rar \Xi^{\prime 0}_b \gamma~~~   & 0.281    & \mbox{---} & 0.047     & \mbox{---}  \\
~~~\Xi^{\ast -}_b     \rar \Xi^{\prime -}_b \gamma      & 0.702    & \mbox{---} & 0.066     & \mbox{---}  \\
~~~\Sigma^{\ast +}_b  \rar \Sigma^+_b       \gamma      & 0.137    &   0.46     & 0.12      & 0.080  \cite{Rvmd15}    \\
~~~\Sigma^{\ast 0}_b  \rar \Sigma^0_b       \gamma      & 0.006    &   0.028    & 0.0076    & 0.005 \cite{Rvmd15}    \\
~~~\Sigma^{\ast -}_b  \rar \Sigma^-_b       \gamma      & 0.040    &   0.11     & 0.03      & 0.020  \cite{Rvmd15}   \\     
~~~\Sigma^{\ast 0}_b  \rar \Lambda^0_b      \gamma      & 221.5    & 114.0      & \mbox{---}& 260.0  \cite{Rvmd15}   \\
~~~\Xi^{\ast 0}_b     \rar \Xi^0_b          \gamma      & 270.8    & 135.0      & \mbox{---}&  \mbox{---}  \\ 
~~~\Xi^{\ast -}_b     \rar \Xi^-_b          \gamma      & 2.246    &   1.5      & \mbox{---}&  \mbox{---}  \\
~~~\Omega^{\ast -}_b  \rar \Omega^-_b       \gamma      & 2.873    & \mbox{---} & 0.00074   &   \mbox{---} \\ \hline \hline
~~~\Xi^{\ast +}_c     \rar \Xi^{\prime +}_c \gamma      & 0.485    & \mbox{---} & 0.96  &  \mbox{---}       \\
~~~\Xi^{\ast 0}_c     \rar \Xi^{\prime 0}_c \gamma      & 1.317    & \mbox{---} & 0.12  &   \mbox{---}      \\
~~~\Sigma^{\ast ++}_c \rar \Sigma^{++}_c    \gamma      & 3.567    &   2.65     & 6.36  &   1.70\cite{Rvmd12},1.15\cite{Rvmd07}     \\
~~~&&&& 3.04 \cite{Rvmd16}\\
~~~\Sigma^{\ast +}_c  \rar \Sigma^+_c       \gamma      & 0.187    &   0.08     & 0.40  & 0.01\cite{Rvmd12},0.00006\cite{Rvmd07}     \\
~~~&&&& 0.19\cite{Rvmd16},0.14\cite{Rvmd13}\\
~~~\Sigma^{\ast 0}_c  \rar \Sigma^0_c       \gamma      & 1.049    &   0.40     & 1.58  &   1.20 \cite{Rvmd12},1.12\cite{Rvmd07}\\
	~~~\Sigma^{\ast +}_c  \rar \Lambda^+_c      \gamma      & 409.3    &   130.0    & \mbox{---}&250.0 \cite{Rvmd12},154.48\cite{Rvmd07}\\
~~~&&&& 151.0 \cite{Rvmd13}, 230.0 \cite{Rvmd14} \\
~~~\Xi^{\ast +}_c     \rar \Xi^+_c          \gamma      & 152.4    &  52.0      & \mbox{---}&  124.0\cite{Rvmd12},63.32\cite{Rvmd07}  \\
~~~&&&& 75.60 \cite{Rvmd14}\\   
~~~\Xi^{\ast 0}_c     \rar \Xi^0_c          \gamma      & 1.318    &   0.66     & \mbox{ ---}&  0.80\cite{Rvmd12},0.30\cite{Rvmd07}  \\
~~~&&&& 0.90 \cite{Rvmd14}\\
~~~\Omega^{\ast 0}_c  \rar \Omega^0_c       \gamma      & 1.439    &  \mbox{---}& 1.16   &0.36\cite{Rvmd12},2.02\cite{Rvmd07}\\
\hline \hline

\end{array}
$$
\caption{Widths of the radiative decays of heavy flavored baryons.}

\renewcommand{\arraystretch}{1}
\addtolength{\arraycolsep}{-1.0pt}
\end{table}

\end{document}